\documentclass[prd,twocolumn,amsmath,superscriptaddress]{revtex4}
\usepackage{amssymb,amsmath}
\usepackage{graphicx}
\usepackage{color}
\usepackage{xfrac}

\usepackage[caption=false]{subfig}

\begin{document}
\title{Boundary conditions and renormalized stress-energy tensor on  a Poincar\'e patch of $\textrm{AdS}_2$}

\author{Jo\~ao Paulo M. Pitelli}
\email[]{pitelli@ime.unicamp.br}
\affiliation{Departamento de Matem\'atica Aplicada, Universidade Estadual de Campinas, 13083-859, Campinas, S\~ao Paulo, Brazil}
\altaffiliation[Also at ]{The Enrico Fermi Institute, The University of Chicago, Chicago, IL}

\author{Vitor S. Barroso}
\email[]{barrosov@ifi.unicamp.br}
\affiliation{Instituto de F\'isica ``Gleb Wataghin'', Universidade Estadual de Campinas, 13083-859, Campinas, S\~ao Paulo, Brazil}

\author{Ricardo A. Mosna}
\email[]{mosna@ime.unicamp.br}
\affiliation{Departamento de Matem\'atica Aplicada, Universidade Estadual de Campinas, 13083-859, Campinas, S\~ao Paulo, Brazil}

\begin{abstract}
Quantum field theory on anti-de Sitter spacetime requires the introduction of boundary conditions at its conformal boundary, due essentially to the absence of global hyperbolicity. Here we calculate the renormalized stress-energy tensor $T_{\mu\nu}$ for a scalar field $\phi$ on the Poincar\'e patch of $\text{AdS}_2$ and study how it depends on those boundary conditions. We show that, except for the Dirichlet and Neumann cases, the boundary conditions break the maximal $\textrm{AdS}$ invariance. As a result, $\langle\phi^2\rangle$ acquires a space dependence and $\langle T_{\mu\nu}\rangle$ is no longer proportional to the metric. When the physical quantities are expanded in a parameter $\beta$ which characterizes the boundary conditions (with $\beta=0$ corresponding to Dirichlet and $\beta=\infty$ corresponding to Neumann), the singularity of the Green's function is entirely subtracted at zeroth order in $\beta$. As a result, the contribution of nontrivial boundary conditions to the stress-energy tensor is free of singular terms.

\end{abstract}

\maketitle

\section {Introduction}

The construction of quantum field theory on curved backgrounds (solutions of Einstein equations) usually assumes that the spacetime is globally hyperbolic (see, for instance, Ref.~\cite{birrel}). This is a very reasonable assumption since, in this case, a Cauchy surface $\Sigma$ exists so that the wave problem is well posed and determined by the initial data at $\Sigma$~\cite{wald0}. However, it is possible to prescribe a sensible evolution for the wave function even when the spacetime is non-globally hyperbolic. This prescription was first presented by Wald~\cite{wald1} and amounts to finding the positive self-adjoint extensions of the spatial part of the wave operator. In~\cite{ishibashi1}, it was shown that this is the only prescription which is consistent with some very reasonable assumptions---essentially related to causality and energy conservation.

It is well known that the anti-de Sitter spacetime is not globally hyperbolic. At the conformal boundary, information can flow in/out from/to infinity so that no Cauchy surface exists in $\textrm{AdS}_n$. In particular, in the Poincar\'e patch $\textrm{PAdS}_n$ given by the metric
\begin{equation}
ds^2=\frac{l^2}{z^2}\left(-dt^2+dz^2+\sum_{i=1}^{n-2}dx_i^2\right),\,\,\,\,z\in(0,\infty),
\label{metric1}
\end{equation}
the conformal boundary is given by $z=0$.

Here, we will treat the case of $\textrm{PAdS}_2$, which will be enough to illustrate our main result: the breaking of $\textrm{AdS}$ invariance and the corresponding extra terms in the stress-energy tensor (which appear in addition to the usual term proportional to the metric). 

The vacuum state depends on the choice of the  boundary condition to be imposed at $z=0$ as follows. A solution $\phi$ of the Klein-Gordon equation,
\begin{equation}
\left(\nabla^\mu\nabla_\mu-m^2-\xi R\right)\phi(x)=0,
\label{wave equation}
\end{equation}
can always be expanded in terms a complete set of normalized modes $u_\omega^{(\beta)}(x)$: 
\begin{equation}
\phi^{(\beta)}(x)=\sum_{\omega}{\left[a_\omega^{(\beta)} u_\omega^{(\beta)}(x)+a_\omega^{(\beta)^\dagger }u_\omega^{(\beta)\ast}(x)\right]}, 
\end{equation}
where mode labels were omitted  for simplicity.
In this equation, $\{u_\omega^{(\beta)}(x)\}$ must satisfy a boundary condition at $z=0$ identified by the parameter $\beta$, which labels possible choices of the self-adjoint extensions (for details, see section II). These modes are eigenfunctions of the Killing vector field $\partial/\partial t$ with eingenvalue $-i \omega$ ($\omega>0$) and are mutually  orthogonal  with respect to the scalar product 
\begin{equation}
\left<\phi_1,\phi_2\right>=-i\int_{0}^{\infty}{\phi_1(x)\overleftrightarrow{\partial_\mu}\phi_2(x)[-g_\Sigma(x)]^{1/2}d\Sigma^{\mu}}.
\end{equation}  
The vacuum state is then given by
\begin{equation}
a_\omega^{(\beta)}|0\rangle_\beta=0\,\,\,\forall \omega.
\end{equation}
As the notation suggests, the construction of the vacuum state crucially depends on the boundary condition and this may be chosen arbitrarily, since each choice of $\beta$ corresponds, in principle, to a legitimate self-adjoint evolution operator for the theory.

We show in this paper that the vacuum  $|0\rangle_\beta$ does not respect $\textrm{AdS}$ invariance unless $\beta$ corresponds to the Dirichlet and Neumann boundary conditions (this fact was already proved for the conformal case by one of the authors in~\cite{pitelli}, by means of a contradiction argument based on Ref.~\cite{allen}). This breaking of invariance results in unusual behavior for several physical quantities. For example, the renormalized square field $\langle\phi^2\rangle_\beta$ turns out to be a function of $z$ notwithstanding the maximal symmetry of the underlying spacetime. Moreover, the renormalized stress-energy tensor $\langle T_{\mu\nu}\rangle_{\text{ren}}$ fails to be proportional to the metric, clearly violating maximal symmetry.

It is known that in the Dirichlet case the two point functions $\langle \varphi(x)\varphi(x')\rangle$ are functions of the geodesic distance $\sigma(x,x')$ \cite{allen}, which is in line with the fact that the  Dirichlet boundary condition respects $\textrm{AdS}$ invariance. By expanding the Green's function in terms of the boundary condition parameter $\beta$,
\begin{equation}
G^{(\beta)}(x,x')=G_0(x,x')+\beta G_1(x,x')+\mathcal{O}(\beta^2),
\label{expansion in beta}
\end{equation} 
we show that the divergence of $G^{(\beta)}(x,x')$ in the limit $x'\to x$ is entirely contained in the Dirichlet contribution $G_0(x,x')$, at least for small $\beta$. The other terms in~(\ref{expansion in beta}) come from the purely analytical interaction at the conformal boundary and are finite in the coincidence limit. Since the formal subtraction was already made in Ref.~\cite{kent} and $\langle\phi^2\rangle_{\beta=0}$ and $\langle T_{\mu\nu}\rangle_{\beta=0}$ are known, we are able to add the contributions due to $\beta$ without any further renormalization.

\section{Boundary conditions}

In two-dimensions, the Poincar\'e patch of AdS is given by the metric
\begin{equation}
ds^2=\frac{1}{z^2}\left(-dt^2+dz^2\right),\,\,\,\,z>0,
\end{equation}
where we set $\Lambda$ such that $l=1$ in Eq.~({\ref{metric1}). The wave equation~(\ref{wave equation}) then becomes 
\begin{align}\label{wave equation2}
\frac{\partial^2\phi(t,z)}{\partial z^2
}-\frac{m_\xi^2}{z^2} \phi(t,z)=\frac{\partial^2 \phi(t,z)}{\partial t^2},\\
\nonumber
\end{align}
with $m_\xi=m^2-2\xi $ (since the scalar curvature in this case is given by $R=-2$).

Note that this equation has the form 
\begin{equation}
\frac{\partial^2 \phi(t,z)}{\partial t^2}=-A\phi(t,z),
\end{equation}
where $A$ is the spatial operator given by (minus) the left-hand side of Eq.~({\ref{wave equation2}}). Since $z>0$, it would seem to be reasonable to set  $C_{0}^{\infty}(0,\infty)$ as the domain of the operator $A$. However, $A$ would not be self-adjoint (even though it is symmetric) in this case. We must accordingly find the positive self-adjoint extensions of $A$ that generate a sensible dynamics for $\phi$. 

This problem was thoroughly analyzed in Ref.~\cite{gitman}, where the authors study the self-adjoint extensions of the operator
\begin{equation}
A=-\frac{d^2}{dz^2}+\frac{\alpha}{z^2},
\end{equation}
with $\alpha\in\mathbb{R}$. If $\alpha\geq3/4$, $A$ is essentially self-adjoint, i.e., it has a unique self-adjoint extension~\cite{simon}, which corresponds to the Dirichlet boundary condition. This case was analyzed in~\cite{kent}. If $-1/4\le\alpha<3/4$, there is an infinite number of self-adjoint extensions for $A$. If $\alpha=-1/4$, only one of these extensions is positive (this corresponds to the Dirichlet boundary condition; all the other choices give rise to bound states). For $-1/4<\alpha<3/4$, there is an infinite number of  positive self-adjoint extensions and a complete orthonormal system of eigenfunctions can be written in terms of a parameter $\beta\geq 0$ and Bessel functions:
\begin{widetext}
\begin{equation}
u_\omega^{(\beta)}(t,z)=\left\{\begin{aligned}&\sqrt{\frac{z}{2}}\frac{J_{\chi}(\omega z)+\gamma(\beta,\omega)J_{-\chi}(\omega z)}{\sqrt{1+2\gamma(\beta,\omega)\cos{(\chi \pi)}+\gamma^2(\beta,\omega)}}e^{-i \omega t},\,\,\,\chi\in(0,1/2)\cup(1/2,\infty),\\
&\frac{1}{\sqrt{\pi \omega}}\frac{\sin{(\omega z)}+\beta(\sfrac{\omega}{\omega_0})\cos{(\omega z)}}{\sqrt{1+\beta^2(\sfrac{\omega}{\omega_0})^2}}e^{-i \omega t},\,\,\,\chi=1/2,
\end{aligned}\right.
\label{normal modes}
\end{equation}
\end{widetext}
with $\chi=\frac{1}{2}\sqrt{1+4\alpha}$ and 
\begin{equation}
\gamma(\beta,\omega)=\beta\frac{\Gamma(1-\chi)}{\Gamma(1+\chi)}\left(\frac{\omega}{2\omega_0}\right)^{2\chi}.
\label{parameter}
\end{equation}
Finally, if $\alpha<-1/4$, there are no self-adjoint extensions with spectre  bounded below, so we will not consider this case. In terms of $\chi$, the range of interest, $-1/4<\alpha<3/4$, corresponds to $0<\chi<1$. Notice that $\chi=1/2$ corresponds to the conformal field with $m=0$ and $\xi=0$.  The momentum parameter $\omega_0$  in Eq.~(\ref{parameter}) was introduced to nondimensionalize $\gamma(\beta,\omega)$. This parameter introduces an energy scale to the problem, which entails the breaking of the $\textrm{AdS}$ invariance of the theory.

\section{Breaking of $\textrm{AdS}$ Invariance}

The aim of this section is to present general arguments as to why one should not expect generic boundary conditions to preserve the AdS symmetry of the theory. A more technical discussion is deferred to the next two sections.

The Killing fields on $\textrm{PAdS}_2$,
\begin{equation}
\begin{aligned}
&\xi_1=\partial_t,\\
&\xi_2=t\partial_t+z\partial_z,\\
&\xi_3=(t^2+z^2)\partial_t+2t z\partial_z,
\end{aligned}
\end{equation}
generate the infinitesimal transformations
\begin{equation}
\begin{array}{lllllllll}
1) \, t & \to & t+\lambda, &
2) \, t & \to & t+\lambda t, &
3) \, t & \to & t+\lambda(t^2+z^2),\\
\phantom{1)} \, z & \to & z, &
\phantom{2)} \, z & \to & z+\lambda z,  &
\phantom{3)} \, z & \to & z+2\lambda zt,
\end{array}
\end{equation}
which clearly preserve the boundary at $z=0$. In this section we 
focus on the conformal case, which corresponds to $\chi=1/2$ in Eq.~(\ref{normal modes}). In this case, the boundary condition determined by the parameter $\beta$ reads 
\begin{equation}
u(t,0)-\frac{\beta}{\omega_0} \frac{\partial u(t,0)}{\partial z}=0,
\label{eq:robinbc}
\end{equation}
so that the Dirichlet and Neumann boundary conditions correspond to $\beta=0$ and $\beta=\infty$, respectively.

The derivative term in Eq.~(\ref{eq:robinbc}) transforms at the conformal boundary as
\begin{equation}
\label{d_transformations}
\begin{array}{lllllll}
1) & \!\!\! \frac{\partial}{\partial z}\to\frac{\partial}{\partial z}, &2)& \!\!\!\frac{\partial}{\partial z}\to (1+\lambda)\frac{\partial}{\partial z},& 3)& \!\!\!\frac{\partial}{\partial z}\to (1+2\lambda t)\frac{\partial}{\partial z}.
\end{array}
\end{equation}
We immediately see that the last two transformations, when applied to the modes $u_{\omega}^{(\beta)}(t,z)$, preserve the form of Eq.~(\ref{eq:robinbc}) only for $\beta=0$ and $\beta=\infty$. We thus see that the AdS symmetry cannot be respected by nontrivial Robin conditions (i.e., those which are neither Dirichlet nor Neumann). 

This breaking of $\textrm{AdS}$ invariance by the boundary conditions affects physical quantities defined by the theory as follows.  Let us first analyze the renormalized quantity $\langle\phi^2\rangle_\beta$. We first notice that, since this quantity is time-independent, we have $\mathcal{L}_{\xi_1}\langle\phi^2\rangle_\beta=0$. However, we already know that the second and third transformations in Eq.~(\ref{d_transformations}) violate AdS invariance, so that we cannot conclude that $\partial_z \langle\phi^2\rangle_\beta=0$. Therefore, $\langle\phi^2\rangle_\beta$ can now depend on $z$ for nontrivial Robin conditions. In the next section we show that this is indeed the case.

Regarding the stress-energy tensor, it is easy to see that if $\mathcal{L}_{\xi_i}\langle T^{\mu}_{\phantom{\mu}\nu}\rangle=0$ for $i=1,2,3$, then one would have $\partial_t \langle T^{\mu}_{\phantom{\mu}\nu}\rangle=0$, $\partial_z \langle T^{\mu}_{\phantom{\mu}\nu}\rangle=0$, $\langle T^{t}_{\phantom{t}t}\rangle=\langle T^{z}_{\phantom{z}z}\rangle$ and $\langle T^{t}_{\phantom{t}z}\rangle=\langle T^{z}_{\phantom{z}t}\rangle=0$, where the last equality follows from time reversal symmetry. However, since $\xi_2$ and $\xi_3$ are no longer symmetries for Robin conditions, we can now only guarantee that $\langle T^{\mu}_{\phantom{\mu}\nu}\rangle$ is independent of time and that $\langle T^{t}_{\phantom{t}z}\rangle=\langle T^{z}_{\phantom{z}t}\rangle=0$. The other components $\langle T^{t}_{\phantom{t}t}\rangle$ and $\langle T^{z}_{\phantom{z}z}\rangle$ might well be different from each other and dependent on $z$. We  show in section V that this is in fact what happens.

\section{Renormalized $\langle \phi^2\rangle_\beta$}

A Green's function can be constructed from Eq.~(\ref{normal modes}) by means of  the mode sum
\begin{equation}
G^{(\beta)}(x,x')\sim \int_0^{\infty}{u_{\omega}^{(\beta)}(t,x)u_{\omega}^{(\beta)}(t',x')d\omega}
\end{equation}
and a regularization. The integral involving the normal modes in Eq.~(\ref{normal modes}) is impractical to be done analytically. However, an expansion of the Green's function up to first order in $\beta$ will be already enough to illustrate our main points about the dependence of physical quantities on the boundary conditions.

We will work with the Hadamard's elementary function given by
\begin{equation}
G^{(\beta)}(x,x')=\mathstrut_\beta\langle 0|\left\{\phi^{\beta)}(x),\phi^{\beta)}(x')\right\}|0\rangle_\beta.
\end{equation}
Up to first order in $\beta$ we have\begin{widetext}
\begin{multline}\label{full green}
G^{(\beta)}(x,x')= \sqrt{zz'}\int_0^{\infty}{J_{\chi}(\omega z)J_{\chi}(\omega z')\cos\omega(t-t')d\omega }\\+\frac{\beta}{2^{2\chi}}\frac{\Gamma(1-\chi)}{\Gamma(1+\chi)} \sqrt{zz'}\int_0^{\infty}{\left(\frac{\omega}{\omega_0}\right)^{2\chi}\left[J_{\chi}(\omega z)J_{-\chi}(\omega z')+J_{-\chi}(\omega z)J_{\chi}(\omega z')-2J_{\chi}(\omega z)J_{\chi}(\omega z')\cos{\pi\chi}\right]\cos\omega(t-t')d\omega }\\+\mathcal{O}(\beta^2). 
\end{multline}
\end{widetext}

The first integral in Eq.~(\ref{full green}) corresponds to the Dirichlet boundary condition ($\beta=0$). It was shown in~\cite{allen} that this term respects all the spacetime symmetries since it is a function of the  spacetime coordinates only through the geodesic distance $\sigma(x,x')$, and it is clearly divergent in the limit $x\to x'$. To calculate the value of $\langle \phi^2\rangle_{\beta=0}$ we can use the Hadamard function for the Dirichlet case which was calculated in~\cite{kent}. Using this result, the value of $\langle \phi^2\rangle_{\beta=0}$ after the Hadamard subtraction is given by
\begin{equation}\begin{aligned}
\langle \phi^2\rangle_{\beta=0}&=\lim_{x'\to x}{\frac{1}{2}\left(G_{\text{Dirichlet}}(\sigma)-G_{H,\text{singular}}(\sigma)\right)}\\&=\frac{\log{2}}{4\pi}-\frac{1}{2\pi}\left[\psi\left(\frac{1}{2}+\chi\right)+\gamma\right],
\end{aligned}
\end{equation}
where $\psi$ is the Digamma function~\cite{abramowitz}, $\gamma$ is the Euler-Mascheroni constant and we set the renormalization scale $M=1$.

Let us now take into account the contribution
of the $\beta$-dependent term in Eq.~(\ref{full green}). In order to do this we take the limit $t'\to t$ and make use of the formula~\cite{glasser}
\begin{align}
&\int_{0}^{\infty}{x^{-s}J_{\mu}(a x)J_{\nu}(b x)dx}=2^{-s}b^{\nu}a^{s-\nu-1}\nonumber
\\&\times \frac{\Gamma((\mu+\nu-s+1)/2)}{\Gamma(\nu+1)\Gamma((\mu-\nu+s+1)/2)}\nonumber
\\&\times {}_2F_1\left(\frac{\nu-\mu-s+1}{2},\frac{\nu+\mu-s+1}{2};\nu+1;\frac{b^2}{a^2}\right),
\end{align}
 with $0<b<a$. Eq.~(\ref{expansion in beta}) then becomes (by considering $z'<z$)
 
 \begin{widetext}
 \begin{multline}
 G^{(\beta)}(z,z')=G_{\text{Dirichlet}}(\sigma)+\frac{\beta}{\omega_0^{2\chi}}\frac{\Gamma(1-\chi)}{\Gamma(1+\chi)}\sqrt{z z'}\Bigg[z'^{-\chi}z^{-\chi-1}\frac{\Gamma\left(\frac{1+2\chi}{2}\right)}{\Gamma(1-\chi)\Gamma\left(\frac{1}{2}\right)} {}_2F_1\left(\frac{1}{2},\frac{1+2\chi}{2};1-\chi;\frac{z'^2}{z^2}\right)
 \\ \hspace{-60pt}+z'^{\chi}z^{-3\chi-1}\frac{\Gamma\left(\frac{1+2\chi}{2}\right)}{\Gamma(1+\chi)\Gamma\left(\frac{1-4\chi}{2}\right)} {}_2F_1\left(\frac{1+4\chi}{2},\frac{1+2\chi}{2};1+\chi;\frac{z'^2}{z^2}\right)
 \\ \hspace{-10pt}-2z'^{\chi}z^{-3\chi-1}\frac{\Gamma\left(\frac{1+4\chi}{2}\right)}{\Gamma(1+\chi)\Gamma\left(\frac{1-2\chi}{2}\right)} {}_2F_1\left(\frac{1+2\chi}{2},\frac{1+4\chi}{2};1+\chi;\frac{z'^2}{z^2}\right)\cos{\pi\chi}\Bigg]+\mathcal{O}(\beta^2). \label{G1}
 \end{multline}
 \end{widetext}

Making the use of the identity~\cite{abramowitz}
\begin{equation}\begin{aligned}
{}_2F_1(a,b;c;z)&=\frac{\Gamma(c)\Gamma(c-a-b)}{\Gamma(c-1)\Gamma(c-b)}\\&\quad\times{}_2F_1(a,b;a+b-c+1,1-z)\\&+(1-z)^{c-a-b}\frac{\Gamma(c)\Gamma(a+b-c)}{\Gamma(a)\Gamma(b)}\\&\quad\times{}_2F_1(c-a,c-b;c-a-b+1,1-z),
\end{aligned}
\end{equation}
we find that the dependence of the Green's function on the boundary condition is given by (up to first order in $\beta$ and in the limit $z'\to z$)
\begin{equation}\begin{aligned}
G_{1}(z)&=\frac{1}{(\omega_0z)^{2\chi}}\left(1+\chi  \frac{\epsilon}{z} \right)\\&\qquad\times\frac{\pi   4^{1-\chi } \sin ^2(\pi  \chi ) \csc (4 \pi  \chi ) \Gamma (1-\chi )}{\Gamma \left(\frac{1}{2}-2 \chi \right) \Gamma \left(\frac{1}{2}-\chi \right) \Gamma (\chi +1)^2},
\end{aligned}
\end{equation}
with $z'=z-\epsilon$. Note that the above expression is clearly nondivergent in the coincidence limit. As a result, the Hadamard subtraction on the full Green's function is given by
\begin{multline}
\label{G11}
\langle \phi^2\rangle_{\beta}=\frac{\log{2}}{4\pi}-\frac{1}{2\pi}\left[\psi\left(\frac{1}{2}+\chi\right)+\gamma\right]\\ \qquad\qquad+\frac{ \beta}{2(\omega_0z)^{2 \chi }}\frac{\pi   4^{1-\chi } \sin ^2(\pi  \chi ) \csc (4 \pi  \chi ) \Gamma (1-\chi )}{\Gamma \left(\frac{1}{2}-2 \chi \right) \Gamma \left(\frac{1}{2}-\chi \right) \Gamma (\chi +1)^2}\\\qquad+\mathcal{O}(\beta^2).
\end{multline}
This explicitly shows that nontrivial boundary conditions ($\beta>0$) break the invariance of the theory. The $\textrm{AdS}$ invariant result is recovered only when $z\to\infty$.

Of particular interest is the conformal case ($\chi=1/2$), which can be analytically solved. In fact, the contribution to the Green's function coming from nonzero values of $\beta$ is given by
\begin{widetext}
\begin{equation}
\begin{aligned}
&G^{(\beta)}(z,z')-G_{\text{Dirichlet}}(\sigma)=\\
&=\frac{\sqrt{zz'}}{\pi}\int_{0}^{\infty}{\left\{\frac{1}{ \left(1+\tilde{\beta} ^2 \omega ^2\right)}\left[\frac{\sin (\omega  z)}{\sqrt{\omega  z}}+\frac{ \tilde{\beta}  \omega  \cos (\omega  z)}{\sqrt{\omega  z}}\right] \left[\frac{\sin (\omega  z')}{\sqrt{\omega  z'}}+\frac{ \tilde{\beta}  \omega  \cos (\omega  z')}{\sqrt{\omega  z'}}\right]-\frac{\sin{(\omega  z)}\sin{(\omega z')}}{\omega}\right\}d\omega},
\end{aligned}
\end{equation}
\end{widetext}
with $\tilde{\beta}=\beta/\omega_0$. This integral can be calculated exactly, and is given by~\cite{gradshteyn}
\begin{equation}
G^{(\beta)}(z,z')-G_{\text{Dirichlet}}(\sigma)=-\frac{e^{\frac{z+z'}{\tilde{\beta} }} \text{Ei}\left(-\frac{z+z'}{\tilde{\beta} }\right)}{\pi },
\end{equation}
where $\text{Ei}(x)$ is the exponential integral function. By using the asymptotic expansion $\text{Ei}(-x)\sim-\frac{e^{-x}}{x}$ we have
\begin{equation}\begin{aligned}
\langle\phi^2\rangle_{\beta}&=\frac{\log{2}}{4\pi}-\frac{e^{\frac{2z}{\tilde{\beta} }} \text{Ei}\left(-\frac{2z}{\tilde{\beta} }\right)}{2\pi }\\&=\frac{\log{2}}{4\pi}+\frac{1}{2\pi\omega_0 z}\beta+\mathcal{O}(\beta^2)
\label{conformal case 2}
\end{aligned}\end{equation}
in the limit of $z'\to z$.
It is easy to see that this exactly agrees with Eq.~(\ref{G11}) in the the limit $\chi\to1/2$. Once again, this $\beta$ dependence shows that the invariance is broken for nontrivial boundary conditions.  

\section{Renormalized Stress-energy Tensor}

To calculate the renormalized stress-energy tensor we use the results of Ref.~\cite{folacci}, where the authors found that, in two dimensions,
\begin{multline}
\langle T_{\mu\nu}\rangle_{\text{ren}}=\frac{1}{2\pi}\Bigg[-w_{\mu\nu}+\frac{1}{2}(1-2\xi)w_{;\mu\nu}
\\ \qquad\qquad\quad+\frac{1}{2}\left(2\xi-\frac{1}{2}\right)g_{\mu\nu}\square w+\xi R_{\mu\nu}w
\\\qquad-g_{\mu\nu}v_1\Bigg]+\Theta_{\mu\nu},
\end{multline}
where
\begin{equation}
\begin{aligned}
& w=\lim_{x'\to x}{W(x,x')}=\lim_{x'\to x}{-2\pi\left(G^{+}(x,x')-G^{+}_{\text{H,sing}}(\sigma)\right)},\\ & w_{\mu\nu}=\lim_{x'\to x}{W(x,x')_{;\mu\nu}},\\
& v_1=-\frac{1}{2}m^2-\frac{1}{2}\left(\xi-\frac{1}{6}\right)R,\\
& \Theta_{\mu\nu}=\frac{\ln{M^2}}{4\pi}\left[-\frac{1}{2}m^2g_{\mu\nu}\right].
\end{aligned}
\label{w}
\end{equation}
The term in $v_1$ is responsible for the trace anomaly and $\Theta_{\mu\nu}$ is a conserved quantity depending on the renormalization scale $M$. Since we want to compare the results for Dirichlet and other Robin boundary conditions, we just set $M=1$.

We start by calculating $T_{tt}$. It follows from the discussion above that
\begin{equation}\begin{aligned}
\langle T_{tt}\rangle_{\text{ren}}=\frac{1}{2\pi}\Bigg[&-w_{tt}+\frac{1}{2}(1-2\xi)w_{;tt}\\&-\frac{1}{2z^2}\left(2\xi-\frac{1}{2}\right)\square w+\frac{\xi w}{z^2}+\frac{v_1}{z^2}\Bigg].
\end{aligned}
\label{result kent}
\end{equation}
The result is known for the Dirichlet case ($\beta=0$) and is given by~\cite{kent}
\begin{equation}\begin{aligned}
\langle T_{tt}\rangle_{\beta=0}=-\frac{1}{8\pi z^2}\Bigg[&\left(-2\chi^2-4\xi+\frac{1}{2}\right)\Bigg(\psi\left(\frac{1}{2}+\chi\right)
\\&\hspace{30pt}+\gamma-\frac{\ln{2}}{2}\Bigg)+\chi^2+\frac{1}{12}\Bigg].
\end{aligned}
\label{result kent2}
\end{equation}
We note that for this case the quantity $w$ in Eq~(\ref{result kent}) is constant since the Dirichlet case is $\textrm{AdS}$ invariant. For nonzero values of $\beta$ this is no longer true since $w$ is then a function of $z$ [see Eq.~(\ref{G1})]. It follows from the wave equation that the $t't'$ derivative of $w$ can be related to its $z'z'$ derivative by 
\begin{equation}
\frac{\partial^2 G_1(x,x')}{\partial t'^2}=\frac{\partial^2 G_1(z,z')}{\partial z'^2}+\frac{(1-4\chi^2)}{4z'^2}G_1(z,z').
\end{equation}

This leads to a renormalized component of $T_{tt}$ of the form
\begin{equation}\begin{aligned}
\langle T_{tt}\rangle_{\beta}-&\langle T_{tt}\rangle_{\beta=0}=\\=\lim_{z'\to z}\Bigg\{&-\frac{\partial^2 G_1(z,z')}{\partial z'^2}-\frac{(1-4\chi^2)}{4z'^2}G_1(z,z')\\&-\frac{1}{z'}\frac{\partial G_1(z,z')}{\partial z'}\Bigg\}+\frac{1}{2z}(1-2\xi)\frac{\partial G_1(z,z)}{\partial z}\\&-\frac{1}{2}(2\xi-1/2)\frac{\partial^2 G_1(z,z)}{\partial z^2}+\frac{\xi}{z^2}G_1(z,z).
\end{aligned}
\end{equation}
As a result, we obtain
\begin{widetext}
\begin{equation}\begin{aligned}
\langle T_{tt}\rangle_{\beta}=&\langle T_{tt}\rangle_{\beta=0}\\&+\frac{\beta}{\omega_0^{2\chi}z^{2+2\chi}}\frac{\pi ^2  2^{-2 \chi -1} (2 \chi -1) (8 \xi  (\chi +1)-2 \chi -1) \sin (\pi  \chi ) \csc (4 \pi  \chi ) }{\chi ^2 (\chi +1) \Gamma \left(\frac{1}{2}-2 \chi \right) \Gamma \left(-\chi -\frac{1}{2}\right) \Gamma (\chi )^3}+\mathcal{O}(\beta^2).
\end{aligned}
\label{Ttt}
\end{equation}
\end{widetext}
We also note that for $\chi=1/2$, i.e., in the conformal case, the first order correction is zero.

Applying the same arguments to calculate $\langle T_{zz}\rangle_\beta$ we find that
\begin{widetext}
\begin{equation}\begin{aligned}
\langle T_{zz}\rangle_{\beta}=&-\frac{1}{8\pi z^2}\Bigg[\left(-2\chi^2-4\xi+\frac{1}{2}\right)\Bigg(\psi\left(\frac{1}{2}+\chi\right)+\gamma-\frac{\ln{2}}{2}\Bigg)+\chi^2+\frac{1}{12}\Bigg]
\\&+\frac{\beta}{\omega_0^{2\chi}z^{2+2\chi}}\frac{  4^{-4 \chi -1} (2 \chi -1) (8 \xi  (\chi +1)-2 \chi -1) \sin (\pi  \chi ) \Gamma (-\chi -1) \Gamma (4 \chi ) }{\chi  \Gamma (\chi )^3}
+\mathcal{O}(\beta^2).
\end{aligned}
\label{Tzz}
\end{equation}
\end{widetext}

The fact that the stress-energy tensor is no longer proportional to the metric for nonzero values of $\beta$ is clearly a manifestation of the loss of AdS invariance of the theory.

We note that $\langle T_{zz}\rangle_\beta$ tends to $\langle T_{zz}\rangle_{\beta=0}$ as $\chi\to 1/2$, as was the case for $T_{tt}$. This is expected on the grounds of Refs.~\cite{romeo,wald2}. In~\cite{romeo},  the authors found the stress-energy tensor in the presence of a single plate which splits the Minkowski spacetime into two disjoint regions. There it was shown that $\langle T_{\mu\nu}\rangle_{\text{ren}}$ depends on the  (Robin) boundary condition at the plate, except for the conformal case. In~\cite{wald2}, it was shown that the renormalized stress-energy tensor for conformal fields in  conformally flat spacetimes is given by
\begin{equation}
\langle T_{\mu\nu}\rangle_{\text{ren}}=\, :\mathrel{\langle T_{\mu\nu}\rangle}:+t_{\mu\nu},
\end{equation}
where $:\mathrel{\langle T_{\mu\nu}\rangle}:$ is the normal ordering operator---which is zero by the results of Ref.~\cite{romeo}---and $t_{\mu\nu}$ is a purely geometrical quantity. In our case  $t_{\mu\nu}$  must be of the form
\begin{equation}
t_{\mu\nu}=\frac{1}{24\pi}g_{\mu\nu},
\end{equation}
due to trace anomaly. It can be easily checked that this is indeed the case in Eq.~(\ref{result kent2}). We note that, since $|0\rangle_{\beta}$ is still invariant under time reversal, $\langle T_{tz}\rangle_{\beta}=\langle T_{zt}\rangle_\beta=0$. 

Finally, we must check whether  our stress-energy tensor is conserved or not. This is  one of Wald's axioms~\cite{wald3} on the construction of $\langle T_{\mu\nu}\rangle$. It is easily seen that the nontrivial component of the divergence of $\langle T_{\mu\nu}\rangle$ is given by
\begin{equation}
\nabla_{\mu}\langle T^{\mu}_{\phantom{\mu}t}\rangle_\beta=-\frac{\left[\langle T^{t}_{\phantom{t}t}\rangle_\beta+\langle T^{z}_{\phantom{z}z}\rangle_\beta\right]}{z}+\frac{\partial \langle T^{z}_{\phantom{z}z}\rangle_\beta}{\partial z}
\end{equation}
and it follows from Eqs.~(\ref{Ttt}) and (\ref{Tzz}) that, in fact,
\begin{equation}
\nabla_{\mu}\langle T^{\mu}_{\phantom{\mu}\nu}\rangle_\beta=0+\mathcal{O}(\beta^2).
\end{equation}

\section{Conclusions}

It is usually assumed that quantum field theory on anti-de Sitter spacetime respects its maximal symmetry. This is due to the fact that the vacua constructed from Dirichlet and Neumann boundary conditions respect AdS invariance. However, the theory says nothing about which boundary condition should be chosen when solving the wave equation. We have shown that if a nontrivial generalized Robin boundary condition (i.e., one which is neither Dirichlet nor Neumann) is used, the maximal AdS symmetry of the theory is broken. This manifests itself in an unexpected behaviour of physical quantities like the stress-energy tensor.

The good news is that the Hadamard decomposition, being essentially geometric, is contained in the symmetric part of the Green's function. This is exactly what we found. The divergence of the Green's function is entirely subtracted at zeroth order, so that the analytic interaction of the field with the conformal boundary does not lead to any further divergences. 

However, there are some strong physical restrictions that must still be respected, regardless the choice of boundary conditions: the divergenceless of $\langle T_{\mu\nu}\rangle$, which states conservation of energy, and trace anomaly, which is purely geometrical in principle. We have shown that this in fact happens.

\acknowledgments

J.~P.~M. Pitelli and V.~S. Barroso and  are grateful to Professor R.~M. Wald and G.~Satishchandran for enlightening discussions and also thank the Enrico Fermi Institute for the kind hospitality. J.~P.~M. Pitelli thanks Funda\c c\~ao de Amparo \`a Pesquisa do Estado de S\~ao Paulo (FAPESP) (Grant No. 2018/01558-9). V.~S. Barroso thanks FAPESP (Grant No. 2018/09575-0). Finally we all thank FAPESP (Grant No. 2013/09357-9).


\begin{thebibliography}{99}
\bibitem{birrel}
N.~D. Birrel and P.~C.~W. Davis, {\it Quantum fields in curved space}, (Cambridge University Press, 1982).

\bibitem{wald0}
 R.~M. Wald, {\it General Relativity}, (University of Chicago Press, Chicago, 1984).

\bibitem{wald1}
R.~M. Wald, {\it Dynamics in nonglobally hyperbolic, static space-times}, J. Math. Phys. {\bf 21}, 2802 (1980).

\bibitem{ishibashi1}
A.~Ishibashi and R.~M. Wald, {\it 
Dynamics in non-globally-hyperbolic static spacetimes II: General analysis of prescriptions for dynamics}, Class. Quant. Grav. {\bf 20}, 3815 (2003).

\bibitem{pitelli}
J. P. M. Pitelli, {\it Comment on ``Hadamard states for a scalar field in anti-de Sitter spacetime
with arbitrary boundary conditions''}, arXiv:1904.10023 [gr-qc].

\bibitem{allen}
B. Allen and T. Jacobson, {\it Vector two-point functions in maximally symmetric spaces}, Comm. Math. Phys. {\bf 103}, 669 (1986).

\bibitem{kent}
C. Kent and E. Winstanley, {\it Hadamard renormalized scalar field theory on anti–de Sitter spacetime}, Phys. Rev. D {\bf 91}, 044044 (2015).


\bibitem{gitman}
D.~M. Gitman, I.~V. Tyutin and  B.~L. Voronov, {\it Self-Adjoint
Extensions and Spectral Analysis in the Calogero Problem}, J. Phys. A: Mathematical and Theoretical  {\bf 43}, 145205 (2010).



\bibitem{simon}
M. Reed and B. Simon, {\it Fourier Analysis, Self-Adjointness} (Academic, New York, 1975).

\bibitem{abramowitz}
M. Abramowitz and I.~A. Stegun, {\it Handbook of Mathematical Functions}, (Washington, DC, 1972).

\bibitem{glasser}
M.~L. Glasser and  E. Montaldi, {\it Some integrals involving Bessel functions}, J. Math. Anal. Appl. {\bf 183 } 577 (1994).

\bibitem{gradshteyn}
I.~S.. Gradshteyn, I.~M. Ryzhik, {\it Table of Integrals, Series and Products}, Academic Press, (2007).

\bibitem{folacci}
Y. D\'ecanini and A. Folacci, {\it Hadamard renormalization of the stress-energy tensor for a quantized scalar field in a general spacetime of arbitrary dimension}, Phys. Rev. D {\bf 78}, 044025 (2008).

\bibitem{romeo}
A. Romeo and A. A. Saharian, {\it Casimir effect for scalar fields under Robin boundary conditions on plates}, J. Phys. A {\bf 35}, 1297 (2002).

\bibitem{wald2}
R.~M. Wald, {\it Axiomatic renormalization of the stress tensor of a conformally invariant field in conformally flat spacetimes}, Annals of Physics {\bf 110}, 472 (1978). 
\bibitem{wald3}
R.~M. Wald, {\it The back reaction effect in particle creation in curved space-time}, Commun.
Math. Phys. {\bf 54}, 1 (1977).
\end{thebibliography}
\end{document}